\documentclass[aps,prb,floatfix,twocolumn,superscriptaddress,showpacs,graphicx]{revtex4-2}
\usepackage{array}[=2016-10-06]
\usepackage{graphicx}
\usepackage{epstopdf}
\usepackage{graphicx}
\usepackage{dcolumn}
\usepackage{bm}
\usepackage[utf8]{inputenc}
\usepackage[T1]{fontenc}
\usepackage{mathptmx}
\usepackage{color}
\usepackage{amsmath}
\usepackage{amssymb}
\usepackage{hologo} 
\usepackage{booktabs} 
\usepackage{tablefootnote} 
\usepackage{float}
\usepackage{xcolor}
\usepackage[normalem]{ulem}
\usepackage[markup=underlined]{changes}

\begin{document}

\title{Design and construction of a cryogenic subcooler-box for supplying single phase supercritical helium to dark matter and gravitational wave experiments}

\author{U. R. Singh}
\affiliation{Deutsches Elektronen-Synchrotron (DESY), Notkestraße 85, D-22607 Hamburg, Germany}

\author{R. Ramalingam}
\affiliation{Deutsches Elektronen-Synchrotron (DESY), Notkestraße 85, D-22607 Hamburg, Germany}

\author{C. Reinhardt}
\affiliation{Deutsches Elektronen-Synchrotron (DESY), Notkestraße 85, D-22607 Hamburg, Germany}

\author{O. Korth}
\affiliation{Deutsches Elektronen-Synchrotron (DESY), Notkestraße 85, D-22607 Hamburg, Germany}

\author{J. Penning}
\affiliation{Deutsches Elektronen-Synchrotron (DESY), Notkestraße 85, D-22607 Hamburg, Germany}

\author{J. Schaffran}
\affiliation{Deutsches Elektronen-Synchrotron (DESY), Notkestraße 85, D-22607 Hamburg, Germany}

\author{A. Lindner}
\affiliation{Deutsches Elektronen-Synchrotron (DESY), Notkestraße 85, D-22607 Hamburg, Germany}

\begin{abstract}
We report on the design, development, and installation of the ALPS Cryo-Platform Subcooler Box (ACPS), which is part of the cryogenic platform being established in the HERA North Hall at DESY to supply helium for cooling large-scale dark-matter and gravitational-wave experiments with very high heat loads. The ACPS is capable of subcooling supercritical helium supplied via the 1.6-km-long HERA transfer line by means of a pipe heat exchanger immersed in a subcooler bath filled with liquid helium produced through Joule–Thomson valves. It is also equipped with numerous cryogenic components, including control valves, flow meters, and safety valves, enabling experimental operation to be carried out directly by the ACPS itself and thereby reducing the cryogenic requirements imposed on the experiments. To support a wide range of experiments, the ACPS provides three transfer lines that deliver different levels of cooling power.
\end{abstract}

\keywords{Cryogenics, heat load, heat exchanger, subcooler box, JT valve, safety valve, and instrumentation and sensors}

\maketitle
\section{Introduction}
Various forthcoming cryogenic experiments in the fields of dark matter and gravitational wave research \cite{Jaeckel2010, Collaboration2019, Calvelli2020, Ballantini2005, Berlin2023, Ahn2024, Aggarwal2021} are planned to be constructed at Deutsches Elektronen-Synchrotron (DESY). These include MADMAX (Magnetized Disc and Mirror Axion Experiment), which focuses on axion dark matter searches \cite{Jaeckel2010, Collaboration2019, Calvelli2020, Calvelli2023, garcia2024, Egge2025}, as well as MAGO (Microwave Apparatus for Gravitational Waves Observation) \cite{Ballantini2005, Berlin2023, Fischer2025} and Cryogenic Mirrors (CMs) for gravitational wave detection \cite{Aggarwal2021, Reinhardt2021, Claudio2024, Franke2024}. Similar to other experiments \cite{Lasenby2020, Khatiwada2021, Ruber2021, Ahn2024, Tang2024}, they will also utilize superconducting magnets \cite{Vande2013, Benda2019, Dallocchio2022, Pontarollo2023, Calvelli2023} and superconducting microwave cavities \cite{Ballantini2005, Fischer2025, Tang2024}, thereby enabling searches for dark matter in the presence of magnetic fields and gravitational waves \cite{Jaeckel2010, Collaboration2019, Aggarwal2021}.

Therefore, these experiments are expected to impose very high heat loads at cryogenic temperatures and become very sensitive to cryogenic conditions, thereby requiring large helium flow rates in various cryogenic phases (superfluid, liquid, and gas) at specific temperatures and as an independent operation at any state. Furthermore, the helium supply to each experiment must carry out independently and without cross-interference. Another important point is to highlight here that these experiments are at preliminary stages, therefore the exact operating boundary condition and heat loads have to be considered with a significant contingency. To meet such requirements, a cryogenic platform is currently under construction in the HERA North Hall (HNH) at DESY. This platform utilizes the existing cryogenic infrastructure from the HERA era \cite{Wolf1988, Horlitz1991, Clausen1991, Meinke1991, Lierl1996, Parker1998, Wanderer2002} recently refurbished to provide helium cooling for ALPS II \cite{Bozhko2019, Bozhko2019HL, Albrecht2021}, and incorporates newly designed and constructed process pipes, distribution boxes, and cryostats. These systems will ultimately be connected to a cryogenic plant on the DESY campus via the 1.6-km-long HERA transfer line (TL) \cite{Wolf1988, Lierl1996, Bozhko2019}. Recent activities have focused on designing and constructing the ALPS Cryo-Platform Sub-cooler Box (ACPS), which will subcool helium after it is supplied through the 1.6-km-long HERA TL to compansate its heat load \cite{Bozhko2019} and provide helium with well-known properties.

In this paper, we first provide a brief overview of the existing cryogenic infrastructure from the HERA period. After that, we present a detailed circuit analysis of the ACPS, designed to support various cryogenic operation modes for the experiments. Additionally, we provide detailed information on how to calculate the mass flow rate for sizing control valves, including Joule–Thomson (JT) valves, pipe heat exchangers (PHEXs) and process pipes, which are used for different experiments. Although the heat loads and operating conditions are only known at a preliminary stage, these calculations can serve as a reference for setting up experiments with similar thermal requirements. Finally, we discuss its 3D design, fabrication, installation and different acceptance tests.

\section{Schematic representation of HERA transfer line, ALPS, ACPS, and other cryogenic systems}
\begin{figure*}[!hbt]
\centering
\includegraphics[angle=0, width=1.0\linewidth]{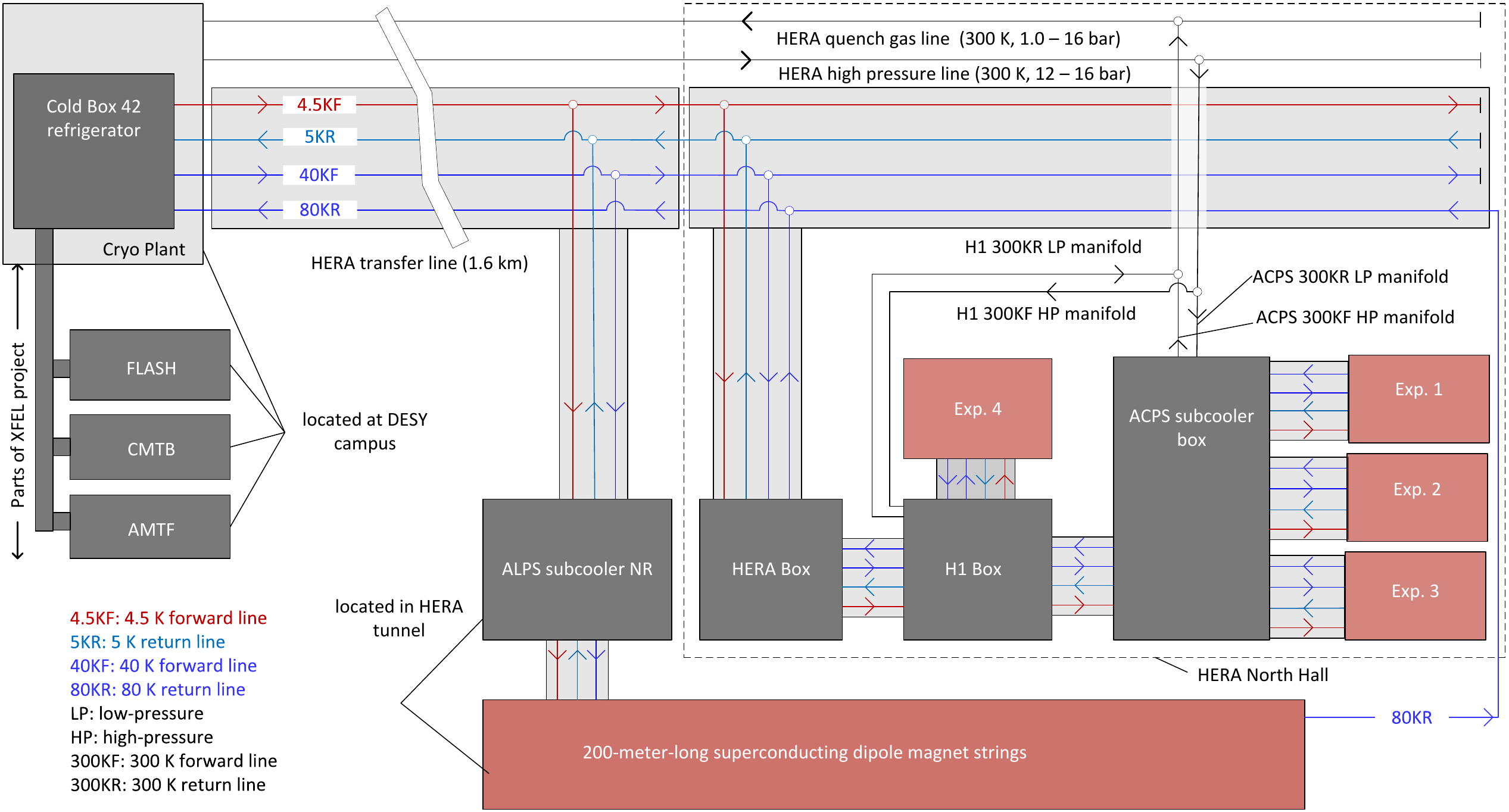}
\caption{Schematic view illustrates the CB42 refrigerator supplying helium to FLASH, AMTF and CMTB (which are test facilities of the XFEL project) at the DESY campus, as well as the HERA TL in the HERA tunnel. At a distance of 1.6 km along the HERA TL, the ALPS subcooler NR and the 200-meter-long dipole magnet string, parts of the ALPS experiment, are connected. Beyond that, the HNH area is located, housing the HERA Box (old), the H1 Box (old), and the newly introduced ACPS with three TL ports designated to three experiments (Exp. 1, 2, and 3). Additionally, the H1 Box includes an option for adding an additional experiment, called Exp.~4, due to its two outlets (as one is used only for the ACPS). The schematic also highlights the warm gas connections.}
\label{fig_HERA_TL_HB_H1_ACPS}
\end{figure*}

Following the shutdown of the HERA accelerator in 2007 \cite{Wolf1988, Horlitz1991, Lierl1996}, the ALPS group at DESY proposed cooling only the 200-meter-long superconducting dipole magnet strings \cite{Albrecht2021} for the second stage of the 'Any Light Particle Search' (ALPS II) experiment \cite{Bähre2013}. To provide helium from the Cold Box 42 (CB42) refrigerator at the DESY campus for this purpose, the MKS group carried out a refurbishment of the existing cryogenic infrastructure from the HERA accelerator, focusing in particular on its performance and capabilities \cite{Bozhko2019, Bozhko2019HL}. This work mainly comprises the HERA TL, the ALPS Subcooler NR, and the HERA Box (located inside the HNH to maintain the 150-meter long extended section of the HERA TL between ALPS and the HNH at cryogenic temperatures \cite{Bozhko2019}), as shown in Fig.~\ref{fig_HERA_TL_HB_H1_ACPS}. The successful operation of the ALPS II experiment in recent years has confirmed that the helium supply provided by these systems works well \cite{Albrecht2021}.

In Fig.~\ref{fig_HERA_TL_HB_H1_ACPS}, we also provide an overview of existing H1 Box \cite{Abt1997} and the newly introduced ACPS, existing in the HNH. It is important to note here that the HERA Box and the H1 Box work as distribution valve boxes, while a state-of-the-art cryogenic system ACPS serves additional as a sub-cooler box, designed to subcool supercritical helium supplied through the 1.6-km-long HERA TL. Furthermore, one can see how and where up to three future experiments (Exp. 1, Exp. 2 and Exp. 3) will be connected to the ACPS and Exp. 4 to the H1 Box.

\subsection{Helium cryogenic lines}
Regarding the different helium cryogenic lines shown in Fig.~\ref{fig_HERA_TL_HB_H1_ACPS}, supercritical helium at 4.5 K and a pressure of 2.5 to 3.5 bar is supplied through the 4.5 K forward (4.5KF) line to various experiments. The return of helium from these experiments, at 4.5-6 K and 1.0-1.3 bar, is handled by the 5 K return (5KR) line. Additionally, the TL includes the 40 K forward (40KF) and 80 K return (80KR) lines, supplying pressurized cold helium at 40-60 K and 14-17 bar and returning it at 60-80 K and 13-16 bar, respectively. (Note: the pressure difference between the 40KF supply and the 80KR return lines at the CB42 should not exceed ~1.0 bar due to defined operating pressure range of CB42). They are also protected by safety valves (SVs) that discharge helium at 20 bar. From this point onwards, the closed circuit formed by the 4.5KF and 5KR lines is referred to as the 4.5–5K circuit, and the circuit formed by the 40KF and 80KR lines is referred to as the 40–80K circuit.

\subsection{300 K high and low pressure manifolds}
In order to carry out pump and purge operations, collect helium during magnet quenches, and mix warm helium with cold helium for cooling down or warming up experiments at a specific temperature gradient, the two warm gas lines are connected to the ACPS and the H1 Box: the 300 K forward (300KF) high-pressure (HP) manifolds and the 300 K return (300KR) low-pressure (LP) manifolds. The 300KF HP manifolds supply pressurized helium gas at 300 K, within a pressure range of 14-17 bar, to the systems. On the other hand, the 300KR LP manifolds collect and return helium from the experiments (via the HERA quench gas line) back to CB42.

\section{Cryogenic operation of ACPS}
\begin{figure*}[!hbt]
\centering
\includegraphics[clip, angle=0, width=1.0\linewidth]{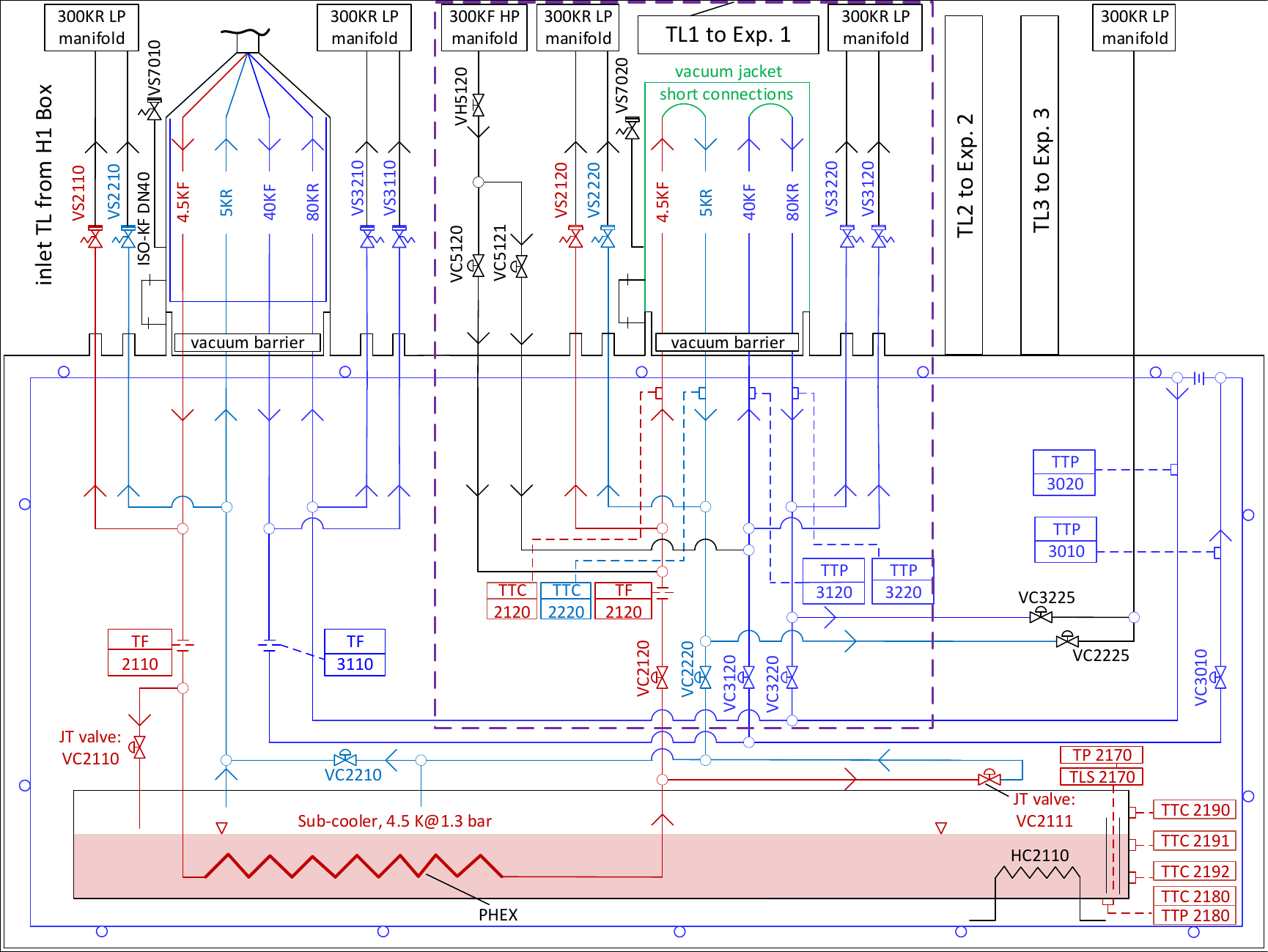}
\caption{An enlarged view of the gray shaded area in Fig.~\ref{fig_ACPS_H1_HB_flow_scheme} (which shows the complete flow scheme of various cryogenic lines and systems in the HNH area) depicts only the inlet TL originating from the H1 Box side, along with a subcooler bath featuring an integrated PHEX and a heater (HC). Additionally, it shows various cryogenic valves (VC), hand valves (VH), safety valves (VS), pressure transmitters (TP), Cernox temperature sensors (TTC), platinum thermometers (TTP), and a flowmeter (TF) for TL1 to Exp. 1. The transfer lines for TL2 to Exp.~2 and TL3 to Exp.~4 are highlighted in Fig.~\ref{fig_ACPS_H1_HB_flow_scheme} in Appendix B.}
\label{fig_ACPS_Box_only_flowscheme}
\end{figure*}

To perform the cryogenic operation of ACPS for subcooling and distribution of helium, we have used various warm and cold valves, a subcooler bath, a PHEX, instrumentation, and sensors. Their functions are described below.

\subsection{JT valves, subcooler bath, and PHEX}
After supercritical helium passes through the 1.6-km-long HERA TL, HERA Box, and H1 Box (see Fig. \ref{fig_HERA_TL_HB_H1_ACPS}), its temperature rises above 4.5 K, primarily due to the heat load onto the HERA TL \cite{Bozhko2019}. To lower the helium temperature back to approximately 4.5 K, the ACPS has been designed as a subcooler box equipped with JT valves and a subcooler bath with an integrated PHEX, as shown in Fig. \ref{fig_ACPS_Box_only_flowscheme} (this is a magnified view of the shaded gray area of ACPS flowscheme presented in Fig. \ref{fig_ACPS_H1_HB_flow_scheme} in Appendix B.). Then, a fraction of the supercritical helium passing through the JT valve VC2110 (VC symbolized as control valve) is liquefied into liquid helium at 4.5 K and 1.3 bar, filling the bath [Fig. \ref{fig_ACPS_Box_only_flowscheme}]. Meanwhile, the rest of helium flows through the PHEX presented in Fig. \ref{fig_ACPS_Box_only_flowscheme} and is subcooled to a temperature close to 4.5 K and an optimal pressure range of 2.5 to 3.5 bar (though in principle, it could go up to 4 to 4.5 bar).

\subsection{40-80K thermal shield}
To minimize parasitic heat transfer from room temperature to all components operating below 6 K or at similar cryogenic temperatures via thermal radiation, structural supports, and other conductive or convective pathways, the ACPS is equipped with a 40–80 K thermal shield, which is conduction-cooled by helium supplied by the HERA TL at 40–80 K and 13–14 bar. The attached cooling pipe to the shield was dimensioned with a diameter of DN20. Helium flow through it is controlled by the VC3010 control valve [see Fig.~\ref{fig_ACPS_Box_only_flowscheme}].

\subsection{TLs to experiments}
As an example, we highlight one of the TLs in the ACPS, namely TL1 to Exp. 1 in Fig. \ref{fig_ACPS_Box_only_flowscheme}, showing various cryogenic lines with the corresponding control valves. It can be seen that both the 4.5KF and 40KF lines are connected to the 300KF HP manifold via VH5120 (VH symbolized as hand valve). This configuration has been designed to facilitate both the cooldown and warm-up of experiments by mixing warm 300 K helium with the cold helium in the 4.5KF and 40KF supply lines. This functionality ensures a controlled temperature gradient, typically in the range of 40–50~K, across the experimental set-up, which helps minimize mechanical stress during cooldown/warm-up and prevent potential damage to the cold mass.

After passing through the experiments during the cooldown and warm-up operations, helium is collected through the 300KR LP manifold, which is further connected to the HERA quench gas line (Fig.~\ref{fig_ACPS_H1_HB_flow_scheme}). In a steady-state operation, it is collected back through 5KR lines, at nearly 4.5 - 6.0 K and 1.1 - 1.3 bar, and returned back to the subcooler bath (which works as a phase-separator collecting liquid helium from the experiments) before sending back it to the CB42 via the HERA TL [see Fig.~\ref{fig_ACPS_Box_only_flowscheme}~and~\ref{fig_ACPS_H1_HB_flow_scheme}]. The 80 KR lines are used for collecting pressurized cold helium at 60-80 K and 13-16 bar from the experiments as well.

\subsection{Safety valves}
Several SVs shown in Fig. \ref{fig_ACPS_Box_only_flowscheme} have been installed on various process pipes to prevent overpressure due to helium expansion in the event of sudden vacuum loss. For example, the 4.5KF and 5KR lines from TL1 to Exp. 1 are protected by VS2120 (VS symbolized as safety valve) and VS2220, while the 40KF and 80KR lines protected by VS3120 and VS3220. SVs for other TL are highlighted in Fig.~\ref{fig_ACPS_H1_HB_flow_scheme}.

\subsection{Instrumentation and sensors}
Fig.~\ref{fig_ACPS_Box_only_flowscheme} shows various instruments and sensors employed throughout the ACPS to monitor and control key parameters during cryogenic processes, including the mass flow rate, He-level, pressure, temperature, and other relevant parameters. We describe their functionalities below.

\textbf{Flowmeters:} To measure the total mass flow of helium into the ACPS, two flowmeters (symbolized as TF) are installed upstream of the box: TF2110 on the 4.5KF line and TF3110 on the 40KF line, as shown in Fig.~\ref{fig_ACPS_Box_only_flowscheme}. Additionally, the 4.5KF line of TL1 to Exp.~1 is equipped with a flowmeter TF2120 (Fig.~\ref{fig_ACPS_Box_only_flowscheme}), while the remaining 4.5KF lines of other TL are equipped with TF2130 and TF2140 (see Fig.~\ref{fig_ACPS_H1_HB_flow_scheme}). For monitoring the mass flow through the 300KF HP manifolds and the 40 KF lines of TL, we monitor the valve positions instead of installing additional flowmeters.

\textbf{Temperature sensors:} To accurately measure the helium temperatures, two types of temperature sensors are installed on various process pipes and the subcooler bath (see Figs.~\ref{fig_ACPS_Box_only_flowscheme} and \ref{fig_ACPS_H1_HB_flow_scheme}). Cernox sensors (TTC) are used to measure temperatures down to 4 K, while platinum resistance thermometers (TTP) are used for temperatures above 40 K. Additionally, TTCs are installed on the subcooler bath (Fig.~\ref{fig_ACPS_Box_only_flowscheme}) at its bottom and at three heights—10\%, 50\%, and 90\%. For redundancy, two additional sensors are added to the primary sensors as backups.

\textbf{Pressure transmitters:} Several pressure transmitters (TP) are installed on the process pipes that handle the helium supply and return at the entrance of the ACPS, as well as on the 300KF HP manifold and the subcooler bath (see Figs.~\ref{fig_ACPS_Box_only_flowscheme} and \ref{fig_ACPS_H1_HB_flow_scheme}). These transmitters are capable of measuring pressures in the range of 0-20 bar.

\textbf{Helium level sensor:} A superconducting helium level sensor, labeled TLS2170 in Fig.~\ref{fig_ACPS_Box_only_flowscheme}, is used to measure the helium level within the subcooler bath.

\textbf{Process heater:} Two heaters, labeled HC2110 in Fig.~\ref{fig_ACPS_Box_only_flowscheme}, are installed on the bottom of the subcooler bath to facilitate the warm-up process and maintain a steady helium flow in the 4.5-5K circuit.

\subsection{Operational Modes of the ACPS}
The operational modes of the ACPS, based on the flow scheme shown in Fig. \ref{fig_ACPS_Box_only_flowscheme}, are outlined below. These modes highlight the system's versatility in meeting various experimental requirements:

\begin{enumerate}
\item Pumping, purging, and leak detection for various cold circuits can be performed effectively and separately for the three experiments.
\item Liquefaction of supercritical helium using JT valves in a subcooler-bath approximately at 4.5 K and a pressure of 1.3 bar
\item Subcooling of supercritical helium to $\thickapprox$ 4.5~$\mathrm{K}$ in a single-phase state using an PHEX immersed in liquid helium, enabling the helium supply at a well-defined and stable temperature.
\item Independent cooldown and warm-up of the three experiments at a controlled temperature gradient can also be performed.
\item Discharge of helium into the quench gas line via various SVs in case of sudden vacuum failure or heat input.
\end{enumerate}

\section{Mass flow rates for ACPS and various experiments}
\begin{figure*}[!hbt]
\centering
\includegraphics[angle=0, width=1.0\linewidth]{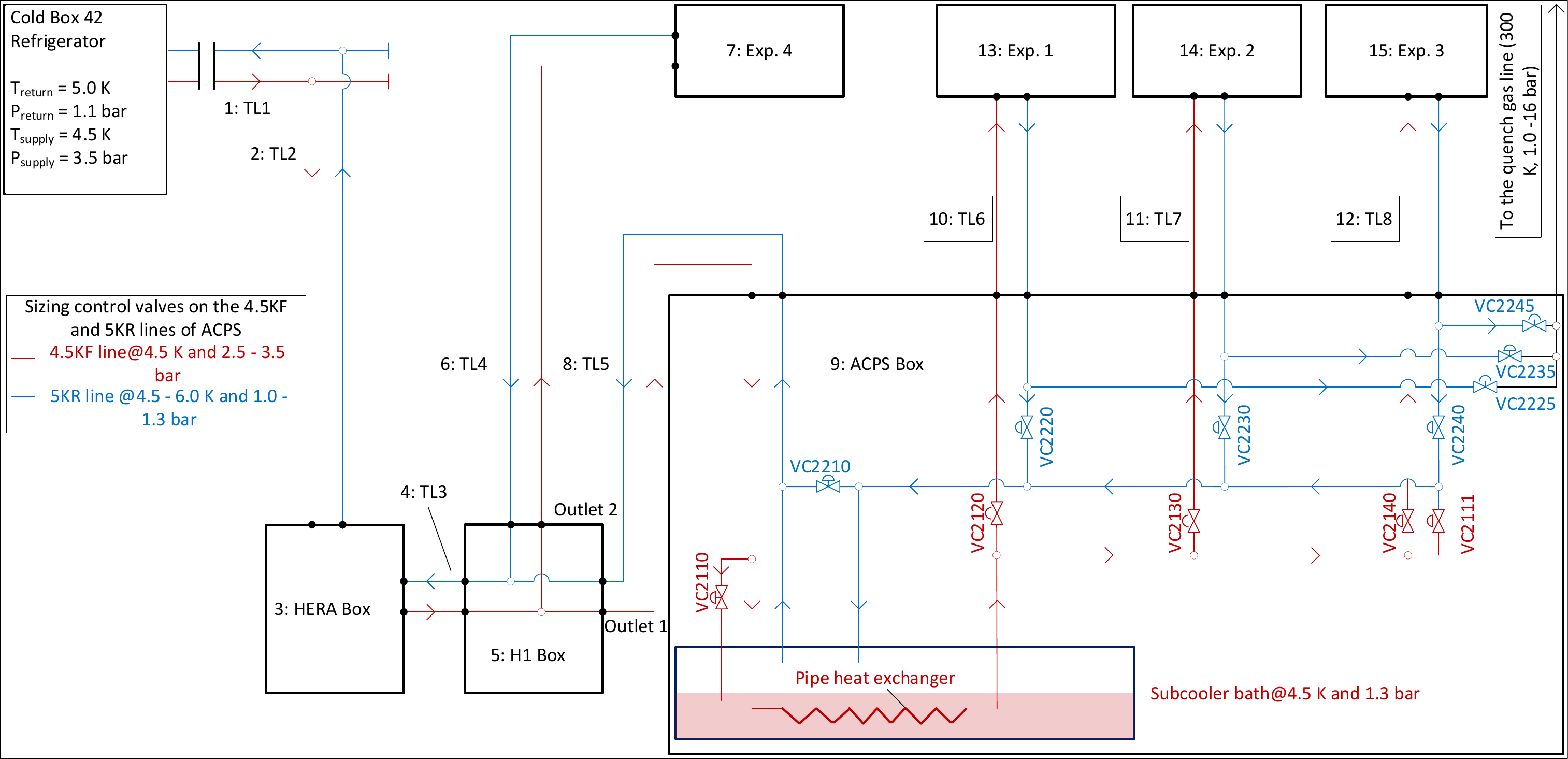}
\caption{4.5-5K circuit diagram of various TLs, distribution boxes, and experiments operated in the temperature range of 4.5 K to 6 K highlighting the heat loads given in Tab.~\ref{tab:heat_loads_sources_4K_table}}.
\label{fig_ACPS_HB_H1_4K_5K_heat_load_v1}
\end{figure*}

\begin{figure*}[!hbt]
\centering
\includegraphics[angle=0, width=1.0\linewidth]{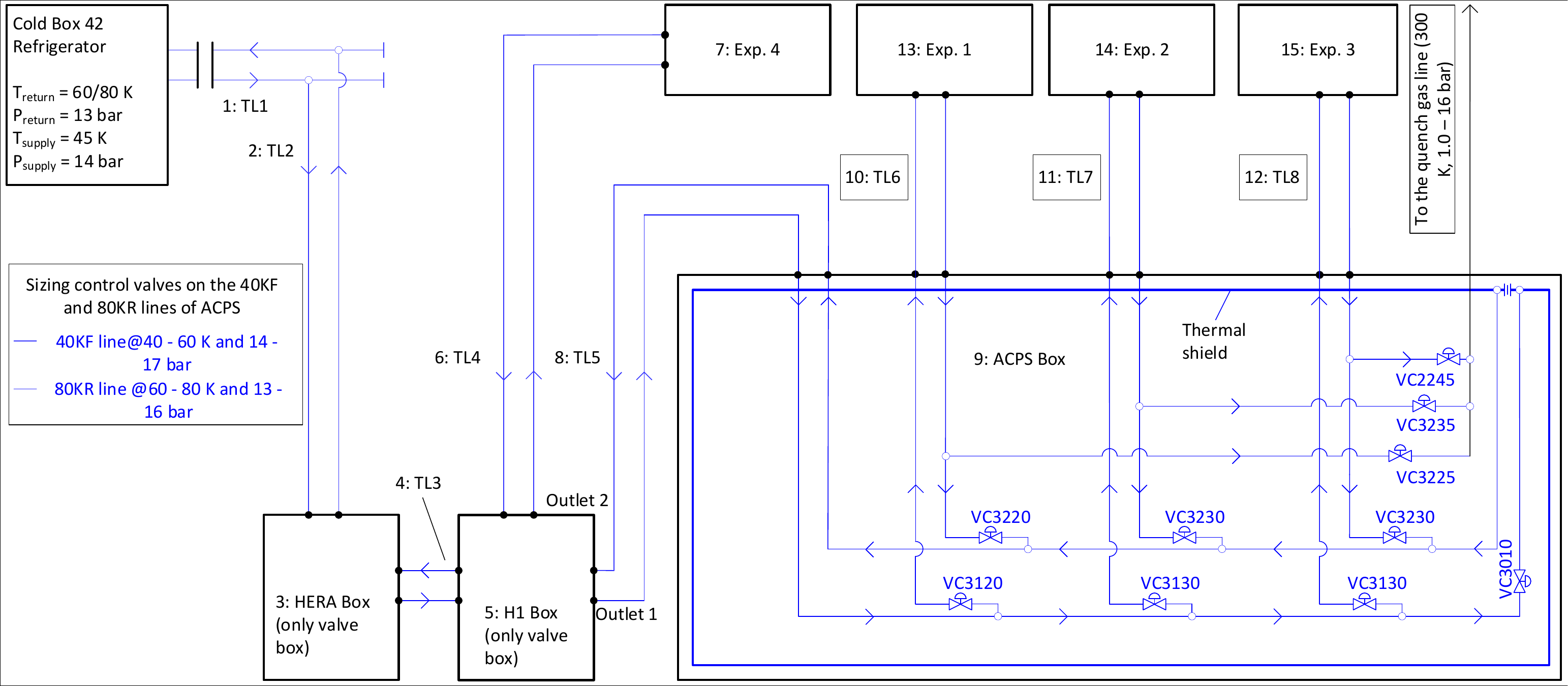}
\caption{40-80K shield circuit diagram of various TLs, distribution boxes, and experiments operated in the temperature range of 40 K to 80 K highlighting the heat loads given in Tab.~\ref{tab:heat_loads_sources_40K_table}}.
\label{fig_ACPS_HB_H1_40_80K_heat_load_v1}
\end{figure*}

The mass flow of helium and the pressure drop through the various cryogenic circuits and systems depend on their heat loads, operating temperatures, and operating pressures. After calculating and knowing these quantities, we could determine the appropriate sizes of valves, pipes, subcooler-bath, mass flow meters, etc. As an overview, we have created circuit diagrams of various TLs, cryogenic boxes, and experiments linked to the 4.5-5K circuit operated in the temperature range of 4.5 K to 6 K shown in Fig. \ref{fig_ACPS_HB_H1_4K_5K_heat_load_v1}, as well as the 40-80K circuit operated at 40-80 K in Fig. \ref{fig_ACPS_HB_H1_40_80K_heat_load_v1}. The corresponding heat load values of different systems in these figures are provided in Tab.  \ref{tab:heat_loads_sources_4K_table} and \ref{tab:heat_loads_sources_40K_table}. Although the future experiments described as Exp. 1-3 in both tables are still at a preliminary stage, their heat loads can be estimated within a range, taking into account a certain contingency. This provides the flexibility to select experiments with comparable heat load requirements.

Using these parameters, one can calculate the required mass flow rates for the cryogenic boxes and experiments under different operation conditions. We potentially consider two conditions for the ACPS:  (a) keeping only the ACPS cold, and (b) keeping the ACPS and all connected experiments cold. The following sections present the mass flow rates for the 4.5-5K and 40-80K circuits of various systems, focusing solely on steady-state operation and all systems being cold (case b here). HEPAK \cite{hepak2021}, a commercial software tool for helium thermophysical properties, is used to obtain enthalpy values and other physical parameters.

\begin{table*}[htbp]
\centering
\caption{\label{tab:heat_loads_sources_4K_table} Summary of heat loads ($\mathrm{HL}_\mathrm{1,~4.5-5K~circuit,~system}$@$(\mathrm{4.5~K})$) to the transfer lines, distribution boxes, and experiments linked to the 4.5-5K circuit shown in Fig. \ref{fig_ACPS_HB_H1_4K_5K_heat_load_v1}. We also highlight Scenario~1 and Scenario~2, representing the minimum and maximum heat loads, respectively.}
\begin{tabular}{p{0.5cm}p{2.5cm}p{3.5cm}p{3.5cm}p{6cm}}
\hline
No. & System & $\mathrm{HL}_\mathrm{1,~4.5-5K~circuit,~system}$ @ $(\mathrm{4.5~K})$ [W] for Scenario~1 & $\mathrm{HL}_\mathrm{2,~4.5-5K~circuit,~system}$ @ $(\mathrm{4.5~K})$ [W] for Scenario~2 & Sources and comments \\
\hline
1 & TL1 & 160 & 320 & Heat load to the 1.6 km-long HERA TL between CB42 and HNH~\cite{ESCHRICHT1988, Clausen1991, Bozhko2019HL}. \\
2 & TL2 & 2.8 & 3.6 & Heat load to the 8 meter-long TL between the HERA TL and the HERA Box~\cite{Bozhko2019, Bozhko2019HL}. \\
3 & HERA Box & 2.0 & 3.2 & Heat load to the HERA Box~\cite{Bozhko2019, Bozhko2019HL}. \\
4 & TL3 & 2.0 & 3.2 & Heat load to the TL between the HERA Box and the H1 Box, based on the HERA TL's per-meter 4 K circuit heat load~\cite{ESCHRICHT1988, Clausen1991, Bozhko2019HL}. \\
5 & H1 Box & 8.65 & 12.1 & Heat load to the H1 Box, including the TL between the H1 Box and the ACPS~\cite{Schaffran2020CP}.\\
6 & TL4 & 4.0 & 6.0 & Heat load to the TL between the H1 Box and Exp.~4, based on the HERA TL's per-meter 4.5-5K circuit heat load~\cite{ESCHRICHT1988, Clausen1991, Bozhko2019HL}. \\
7 & Exp. 4$^{a}$ & 40.0 & 60.0 & Heat load to Exp.~4, assumed for the MADMAX booster cryostat~\cite{Schaffran2020CP}, introduced as a potential experiment~\cite{Collaboration2019, garcia2024, Egge2025}. \\
8 & TL5 & --- & --- & Already included in the heat load value of the H1 Box. \\
9 & ACPS & 36.4 & 58.38 & Heat load to the ACPS box. \\
10 & TL6, TL7, TL8 & 3$\times$4.0 & 3$\times$6.0 & Heat load to the TL between the ACPS and Exp.~1, based on the HERA TL's per-meter 4.5-5K circuit heat load~\cite{ESCHRICHT1988, Clausen1991, Bozhko2019HL}. \\
11 & Exp. 1$^{b}$ & 117 & 177.5 & Heat load to Exp.~1, assumed for the MADMAX experiment~\cite{Collaboration2019, Calvelli2020, Calvelli2023, garcia2024, Schaffran2020CP}, including a helium liquefaction rate of $\dot{m}_\mathrm{LR,~Exp.~1} = 3.0~\mathrm{g/s}$ for the 1.8 K bath~\cite{Schaffran2020CP}. \\
12 & Exp. 2$^{c}$ & 20 & 25 & Heat load to Exp.~2, assumed for the CMs experiment~\cite{Reinhardt2021, Claudio2024, Schaffran2020CP}, with a helium liquefaction rate of $\dot{m}_\mathrm{LR,~Exp.~2} = 3.0~\mathrm{g/s}$ (1.8 K bath)~\cite{Schaffran2020CP}. \\
13 & Exp. 3$^{d}$ & 10 & 12.5 & Heat load to Exp.~3, assumed for the MAGO~\cite{Ballantini2005, Berlin2023, Fischer2025} or COMAG experiment~\cite{Schaffran2020CP}, with a helium liquefaction rate of $\dot{m}_\mathrm{LR,~Exp.~3} = 0.18~\mathrm{g/s}$~\cite{Schaffran2020CP}. \\
\hline
\hline
\end{tabular}
\footnotetext{$a, b, c, d$: The heat loads for these experiments are determined at a preliminary stage, but can be used as a reference for setting up experiments with similar heat loads.}
\end{table*}

\begin{table*}[htbp]
\centering
\caption{\label{tab:heat_loads_sources_40K_table} Summary of heat loads ($\mathrm{HL}_\mathrm{1,~40-80K~circuit,~system}$@$(\mathrm{40~K})$) to the 40-80K shield circuit of various TL, distribution boxes, and experiments shown in Fig. \ref{fig_ACPS_HB_H1_40_80K_heat_load_v1}.}
\begin{tabular}{p{0.5cm}p{2.5cm}p{3.5cm}p{3.5cm}p{6cm}}
\hline
No. & System & $\mathrm{HL}_\mathrm{1,~40-80K~circuit,~system}$ @ $(\mathrm{40~K})$ (W) for Scenario~1 & $\mathrm{HL}_\mathrm{2,\,~40-80K~circuit,~system}$ @ $(\mathrm{40~K})$ (W) for Scenario~2 & Sources and comments \\
\hline
1 & TL1 & 960 & 2100 & Heat load to the 1.6 km-long HERA TL between CB42 and HNH~\cite{ESCHRICHT1988, Clausen1991, Bozhko2019HL}. \\
2 & TL2 & 7.5 & 15  & Heat load to the 8-meter-long TL between the HERA TL and the HERA Box~\cite{Bozhko2019, Bozhko2019HL}. \\
3 & HERA Box & 8.4 & 14.4 & Heat load to the HERA Box~\cite{Bozhko2019, Bozhko2019HL}. \\
4 & TL3 & 7.5 & 15 & Heat load to the TL between the HERA Box and the H1 Box, based on the per-meter 40 K circuit heat load of the HERA TL~\cite{ESCHRICHT1988, Clausen1991, Bozhko2019HL}. \\
5 & H1 Box & 13.1 & 21.79 & Heat load to the H1 Box, including its thermal shield and the TL between the H1 Box and the ACPS~\cite{Schaffran2020CP}. \\
6 & TL4 & 16.0 & 32.0 & Heat load to the TL between the H1 Box and Exp. 4, based on the per-meter 40 K circuit heat load of the HERA TL ~\cite{ESCHRICHT1988, Clausen1991, Bozhko2019, Bozhko2019HL}. \\
7 & Exp. 4$^{a}$ & 500.0 & 600.0 & Heat load to Exp. 4, assumed for the MADMAX booster cryostat~\cite{Schaffran2020CP}. \\
8 & TL5 & --- & --- & Already included in the heat load value of the H1 Box. \\
9 & ACPS & 36.4 & 58.38 & Heat load to the 40-80K components of the ACPS and its thermal shield. \\
10 & TL6, TL7, TL8 & 3$\times$16.0 & 3$\times$32.0 & Heat load to the TL between the ACPS and Exp. 1, based on the per-meter 40 K circuit heat load of the HERA TL. \\
11 & Exp. 1$^{b}$ & 3700 & 4000 & Heat load to Exp. 1, assumed for the MADMAX experiment~\cite{Collaboration2019, Calvelli2020, Calvelli2023, garcia2024, Schaffran2020CP}, including current leads~\cite{Schaffran2020CP}. \\
12 & Exp. 2$^{c}$ & 150 & 187.5 & Heat load to Exp. 2, assumed for the CMs experiment~\cite{Reinhardt2021, Claudio2024, Schaffran2020CP}. \\
13 & Exp. 3$^{d}$ & 30 & 37.5 & Heat load to Exp. 3, assumed for the MAGO~\cite{Ballantini2005, Berlin2023, Fischer2025} or COMAG experiment~\cite{Schaffran2020CP}. \\

\hline
\multicolumn{2}{l}{\textbf{Total}} & 5476.9 &  7177.6 & \\
\hline
\end{tabular}
\footnotetext{$a, b, c, d$: The heat loads to the 40-80K shield circuit of these experiments are determined at a preliminary stage, but can be used as a reference for setting up experiments with similar heat loads.}
\end{table*}

\subsection{Mass flow rates for the systems connected to the 4.5-5K circuits}
The heat loads to various TLs and systems operating at 4.5 K, whose circuit diagram is shown in Fig.~\ref{fig_ACPS_HB_H1_4K_5K_heat_load_v1}, are listed in Tab.~\ref{tab:heat_loads_sources_4K_table}. It should be noted that helium flowing through the 4.5KF lines of various TLs and boxes experiences a temperature rise before reaching the ACPS due to the heat loads to these systems. To account for this temperature increase, the total heat load to the 4.5–5K circuit is distributed between the 4.5KF and 5KR lines according to their surface-area ratios. For instance, in the HERA TL, the surface-area ratio between these two lines is 25:100 \cite{Bozhko2019, Bozhko2019HL}. However, a simplified pessimistic 50:50 division is adopted to provide a sufficient contingency for the 4.5KF forward line. Furthermore, two scenarios are considered: (1) Scenario 1 represents the minimum heat load (see column 3 in Tab.~\ref{tab:heat_loads_sources_4K_table}), and (2) Scenario 2 represents the maximum heat load (see column 4 in Tab.~\ref{tab:heat_loads_sources_4K_table}).

Thus, to determine the minimum heat load to the 4.5KF line for Scenario 1, $\mathrm{HL}_\mathrm{1,~4.5KF,~total}$, we add half of heat loads to the transfer lines and boxes given in rows 1 to 10 of column 3 of Tab.~\ref{tab:heat_loads_sources_4K_table}, together with the heat loads from the individual experiments listed in the same column. The result is $\mathrm{HL}_\mathrm{1,~4.5KF,~total} = 292.65~\mathrm{W}$.
Considering the enthalpy of helium supply from the CB42 refrigerator, $h_\mathrm{supply,~4.5KF,~CB42}@(\mathrm{4.5~K,~3.5~bar}) = 11.92~\mathrm{J/g}$~\cite{hepak2021}, and the enthalpy of helium return to the ACPS, $h_\mathrm{return,~5KR,~ACPS}@(\mathrm{4.5~K,~1.3~bar}) = 30.5~\mathrm{J/g}$~\cite{hepak2021}, the corresponding required minimum mass flow is $\dot{m}_\mathrm{1,~4.5KF,~total} = 15.75~\mathrm{g/s}$.

Similarly, for Scenario 2, the maximum heat load to the 4.5KF line is calculated by considering the heat load values given in column 4 of Tab.~\ref{tab:heat_loads_sources_4K_table}, $\mathrm{HL}_\mathrm{2,~4.5KF,~total} = 475.9~\mathrm{W}$. After adding the total liquefaction rate for Exp.~1, Exp.~2, and Exp.~3, $\dot{m}_\mathrm{LR,~Exps.~1,~2,~3} = 6.18~\mathrm{g/s}$ \cite{Schaffran2020CP}, the maximum mass flow rate can be $\dot{m}_\mathrm{2,~4.5KF,~total} = 31.79~\mathrm{g/s}$. Thus, $\dot{m}_\mathrm{1,~4.5KF,~total}$ and $\dot{m}_\mathrm{2,~4.5KF,~total}$ show the helium mass flow required for Exp.~1~-~3 at 4.5 K in the range of 16-32 g/s.

\subsubsection{Mass flow rate for Exp. 4}
Since a fraction of helium is extracted for Exp.~4 [see Fig.~\ref{fig_ACPS_HB_H1_4K_5K_heat_load_v1}], which is connected to the H1 Box, this mass flow must be known before determining the helium temperature at the ACPS. In this case, Exp.~4 will not be supplied with subcooled helium; therefore, the helium temperature will be higher due to the high heat load to the HERA TL. To calculate the mass flow rate, the enthalpy of helium at the H1 Box for both scenarios must first be determined. (Note that the calculation is shown in this paper only for Scenario~1, the same applies to the other calculations described in the following subsections.). The total heat load to the 4.5KF line from CB42 up to the H1 Box (see the 4.5KF line in Fig.~\ref{fig_ACPS_HB_H1_4K_5K_heat_load_v1}) is obtained by summing the heat load of Systems 1~-~5 in Tab.~\ref{tab:heat_loads_sources_4K_table} and dividing it by 2. For Scenario 1, this results in $\mathrm{HL}_{\mathrm{1,~4.5KF,~CB42-H1~Box}} = 87.72~\mathrm{W}$. Now, the inlet enthalpy of helium supply at the H1 box to Exp.~4 is calculated as follows,
\begin{multline*}
h_{\mathrm{1,~4.5KF,~H1~Box}} = \frac{\mathrm{HL}_{\mathrm{1,~4.5KF,~CB42-H1~Box}}}{\dot{m}_{\mathrm{1,~4.5KF,~total}}} \\
+ h_{\mathrm{supply,~4.5KF,~CB42}} = 17.48~\mathrm{J/g},
\end{multline*}
yielding the helium temperature, $T_{\mathrm{1,~4.5KF,~H1~Box}} = 5.40~\mathrm{K}$~\cite{hepak2021}.

Thus, the required mass flow rate for a steady state operation of Exp.~4, based on $h_{\mathrm{1,~4.5KF,~H1~Box}}$, the outlet enthalpy of helium return to the H1 Box ($h_\mathrm{return,~5KR,~H1~Box}@(\mathrm{4.5~K,~1.3~bar}) = 30.5~\mathrm{J/g}$), along with heat loads given for TL4 and Exp.4 in rows 6 (this will be divided by 2 for the 4.5KF line only) and 7 of column 3 of Tab.~\ref{tab:heat_loads_sources_4K_table}, respectively, is obtained to be $\dot{m}_{\mathrm{1,~4.KF,~Exp.~4}} = 3.22~\mathrm{g/s}$. This mass flow gets extracted for Exp.~4, prior to helium being supplied to the ACPS and the remaining experiments.

\subsubsection{Heat transfer by the PHEX and evaporation rate of helium in subcooler bath}
As shown in Fig.~\ref{fig_ACPS_HB_H1_4K_5K_heat_load_v1}, the ACPS is equipped with an PHEX to sub-cool helium supplied via the HERA TL. For designing the PHEX, the inlet and outlet temperatures as well as the required mass flows must be defined. Therefore, we first estimate the enthalpy of helium for Scenario~1 at the ACPS using $\mathrm{HL_{1,~4.5-5K~circuit,~ACPS}}$ given for ACPS in row 9 of column 3 of Tab.~\ref{tab:heat_loads_sources_4K_table}:
\begin{multline*}
h_\mathrm{1,~4.5KF,~ACPS} = h_{\mathrm{1,~4.5KF,~H1~Box}} + \\
\frac{\mathrm{HL_{1,~4.5-5K~circuit,~ACPS}/2}}{\dot{m}_\mathrm{1,~4.5KF,~total} - \dot{m}_{\mathrm{1,~4.5KF,~Exp.~4}}} = 18.28~\mathrm{J/g}.
\end{multline*}
This leads to the inlet temperature at the PHEX for Scenario~1, $T_{\mathrm{1,~4.5KF,~inlet-HeX}} = 5.48~\mathrm{K}$~\cite{hepak2021}, being higher than the temperature of helium supply from CB42, $T_\mathrm{supply,~4.5KF,~CB42}~\approx~4.5~\mathrm{K}$. Therefore, the helium must be subcooled to $T_\mathrm{supply,~4.5KF,~CB42}$ using the PHEX before being distributed to various experiments.

For this, we estimate the heat transfer ($\dot{Q}_{\mathrm{1,~4.5KF,~PHeX}}$) by the PHEX to liquid helium in the subcooler bath as follows:
\begin{multline*}
\dot{Q}_{\mathrm{1,~4.5KF,~PHeX}} = (\dot{m}_{\mathrm{1,~4.5KF,~total}} - \dot{m}_{\mathrm{1,~4.5KF,~Exp.~4}}) \times \\
(h_\mathrm{1,~4.5KF,~ACPS} - h_{\mathrm{supply,~4.5KF,~CB42}}) = 79.67~\mathrm{W}. \\
\end{multline*}
Using this value, the evaporation rate of helium from the subcooler bath is calculated as,
\begin{multline*}
\dot{m}_{\mathrm{1,~4.5KF,~subcooler}} = \frac{\dot{Q}_{\mathrm{1,~4.5KF,~PHeX}}}{L_\mathrm{fg}}, \\
\end{multline*}
where $L_\mathrm{fg}$ is the latent heat of helium, and its value is $18.86~\mathrm{J/g}$@($4.5~\mathrm{K},~1.3~\mathrm{bar}$)~\cite{hepak2021}. In this way, we find $\dot{m}_{\mathrm{1,~4.5KF,~subcooler}} = 4.24~\mathrm{g/s}$, which has to be maintained by liquefying helium via the JT valves either VC2110 or VC2111, as shown in Fig. \ref{fig_ACPS_HB_H1_4K_5K_heat_load_v1}.

\subsubsection{Liquefaction rate via JT valve VC2110}
For a continuous production of $\dot{m}_{\mathrm{1,~4.5KF,~subcooler}}$ via the JT valve VC2110 into subcooler bath shown in Fig.~\ref{fig_ACPS_HB_H1_4K_5K_heat_load_v1}, we need to estimate its initial mass flow rate using the yield factor, $x$, based on equations described in Appendix C. This requires the inlet enthalpy value of helium entering the JT valve, $h_\mathrm{1,~4.5KF,~ACPS}$ (as described above), as well as the enthalpy values of saturated liquid ($h_\mathrm{sat.~liquid,~subcooler}$)@(4.5 K, 1.3 bar) and gaseous helium ($h_\mathrm{sat.~gas,~subcooler}$)@(4.5 K, 1.3 bar), which are taken from Ref.~\cite{hepak2021}. Now, $x$ for Scenario 1 is calculated as follows:
\begin{multline*}
x_\mathrm{1,~4.5KF,~VC2110} = \frac{h_\mathrm{1,~4.5KF,~ACPS} - h_\mathrm{sat.~gas,~subcooler}}{h_\mathrm{sat.~liquid,~subcooler} - h_\mathrm{sat.~gas,~subcooler}} \\
= \frac{18.28 - 30.50}{11.64 - 30.50} = 0.65.
\end{multline*}
Finally, the required inlet helium flow rates via VC2110 is
\begin{multline*}
\dot{m}_\mathrm{1,~4.5KF,~VC2110} = \frac{\dot{m}_\mathrm{1,~4.5KF,~subcooler}}{x_\mathrm{1,~4.5KF,~VC2110}} = 6.52~\mathrm{g/s}.
\end{multline*}

\subsubsection{Liquification rate by JT Valve VC2111}
Using the same methodology as for the VC2110 JT valve, we calculate the initial mass flow for VC2111 ($\dot{m}_\mathrm{1,~4.5KF,~VC2111}$), located downstream of the PHEX in Fig.~\ref{fig_ACPS_HB_H1_4K_5K_heat_load_v1}. When helium enters VC2111, its temperature is already about 4.5~K, therefore, its inlet enthalpy at VC2111 is $h_\mathrm{1,~4.5KF,~VC2111}@(\mathrm{4.5~K},~3.5~\mathrm{bar}) = 11.92$, taken from Ref.~\cite{hepak2021}. Under these conditions, a high yield factor of $x_\mathrm{1,~4.5KF,~VC2111} = 0.98$ is achieved and the mass flow of helium through VC2111 is $\dot{m}_\mathrm{1,~4.5KF,~VC2111} = 5.3~\mathrm{g/s}$.

\subsubsection{Mass flow rate for Exp. 1}
The helium after passing through the ACPS is already subcooled and therefore being supplied at approximately 4.5~K and 3.5~bar to the experiments. This results into an inlet enthalpy of helium supply from the ACPS to Exp.1, $h_\mathrm{supply,~4.5KF,~ACPS}@(\mathrm{4.5~K,~3.5~bar}) = 11.92~\mathrm{J/g}$. The outlet enthalpy of helium return from Exp. 1 to the ACPS is expected to be $h_\mathrm{return,~5KR,~ACPS}@(\mathrm{4.5~K,~1.3~bar}) = 30.5~\mathrm{J/g}$. The required mass flow rates for Scenario~1 of Exp. 1, based on the heat loads given for TL6 and Exp.~1 in rows 10 (this will be divided
by 2 for the 4.5KF line only) and 11 of column 3 of Tab.~\ref{tab:heat_loads_sources_4K_table}, respectively, is obtained to be $\dot{m}_{\mathrm{1,~4.5KF,~Exp.~1}} = 6.40~\mathrm{g/s}$. Similarly, for the heat load values given in Tab.~\ref{tab:heat_loads_sources_4K_table} for Scenario~2, and taking into account the liquefaction rate ($\dot{m}_\mathrm{LR\text{-}Exp.~1} = 3~\mathrm{g/s}$)~\cite{Schaffran2020CP}, the mass flow rate becomes, $\dot{m}_{\mathrm{2,~4.5KF,~Exp.~1}} = 12.71~\mathrm{g/s}$.

To obtain the required mass flow rates for Exp.~2 and Exp.~3, we consider the heat load values given in rows 10, 12, and 13 of Tab.~\ref{tab:heat_loads_sources_4K_table}, as well as the same inlet and outlet enthalpies used for Exp.~1. We summarize all the mass flow rates at 4.5~K described for different JT valves and experiments in Tab.~\ref{tab:mass_flow_rates_4.5KF}.

\begin{table*}[htbp]
\centering
\caption{\label{tab:mass_flow_rates_4.5KF} Summary of the calculated mass flow rates for various experiments shown in Fig.~\ref{fig_ACPS_HB_H1_4K_5K_heat_load_v1} depending on their heat loads given in Tab.~\ref{tab:heat_loads_sources_4K_table} and different enthalpy values of the helium supply and return.}
\begin{tabular}{p{2cm}p{1.5cm}p{1.5cm}p{1.5cm}p{2cm}p{2cm}p{2cm}p{1.5cm}p{1.5cm}p{1.5cm}}
\hline
Cases & $\dot{m}_\mathrm{i,~j,~total}$ (g/s) & $\dot{m}_\mathrm{i,~j,~Exp.~4}$ (g/s) & $\dot{m}_\mathrm{i,~j,~VC2110}$ (g/s) & $\dot{m}_\mathrm{i,~j,~VC2111}$ (g/s) & $\dot{m}_\mathrm{i,~j,~Exp.~1}$ (g/s) & $\dot{m}_\mathrm{i,~j,~Exp.~2}$ (g/s) & $\dot{m}_\mathrm{i,~j,~Exp.~3}$ (g/s)\\
\hline
1: Scenario 1 & 15.75 & 3.22 & 6.52 & 5.33 & 6.40 & 1.18 & 0.68 \\
2: Scenario 2 & 31.79 & 4.77 & 13.10 & 8.83 & 12.71 & 4.50 & 1.01 \\
\hline
\end{tabular}
\footnotetext{Here, $i = 1~\mathrm{and}~2$ representing various cases as well as $j$ denoting 4.5KF}
\end{table*}

\subsection{Mass flow rates for the systems connected to the 40-80K circuits}
The heat loads to various systems including thermal shields connected to the 40-80K circuits, depicted in Fig.~\ref{fig_ACPS_HB_H1_40_80K_heat_load_v1}, are summarized in Tab.~\ref{tab:heat_loads_sources_40K_table}. For the mass flow rate calculation, we assume that the temperature of helium supply from the CB42 via the 40KF line can be $\approx$ 45 K and the pressure is $\approx$ 14 bar, while the temperature of the helium return via the 80KR line to the CB42 can be between 60~K (return1) and 80~K (return2), as well as the pressure is $\approx$  13~bar (A total pressure drop of approximately 1 bar along the entire 40-80K circuit is taken into account, since the inlet pressure of CB42  has to be $\approx$ 13 bar, see Fig.~\ref{fig_ACPS_HB_H1_40_80K_heat_load_v1}). Therefore, we consider the enthalpy values at these temperatures and pressures obtained from Ref. \cite{hepak2021}: $h_\mathrm{supply,~40KF,~CB42}@(\mathrm{45~K,~14~bar}) = 249.0~\mathrm{J/g}$, $h_\mathrm{return1,~80KR,~CB42}@(\mathrm{60~K,~13~bar}) = 327.1~\mathrm{J/g}$ and $h_\mathrm{return2,~80KR,~CB42}@(\mathrm{80~K,~13~bar}) = 431.1~\mathrm{J/g}$. Combining these two return temperature conditions with both Scenario 1 and 2 result into a total of four different cases for each experiment.

As an example, we highlight the mass flow rate calculation for helium return at 60~K for Scenario 1 (referred to as Case 1) for Exp. 4 and Exp. 1. We first determine the total mass flow rate ($\dot{m}_\mathrm{1,~40-80K~circuit,~total}$) for the 40-80K circuit, using $\mathrm{HL}_\mathrm{1,~40-80K~circuit,~total} = 5476.9 \mathrm{W}$ (see the last row of column 3 of Tab.~\ref{tab:heat_loads_sources_40K_table}), $h_\mathrm{supply,~40KF,~CB42}$, and $h_\mathrm{return1,~60KR,~CB42}$, $
\dot{m}_\mathrm{1,~40-80K~circuit,~total} = 70.12~\mathrm{g/s}$, which is distributed across four different experiments.

To determine the mass flow rate for Exp.~4, we first calculate the inlet enthalpy of the helium ($h_\mathrm{1,~40KF,~H1~Box}$) supplied via the 40KF line of the H1 box to Exp.~4 using the 40-80K circuit shown in Fig.~\ref{fig_ACPS_HB_H1_40_80K_heat_load_v1} and by estimating the sum of the heat loads to the systems corresponding to rows 1~-~6 in column 3 of Tab.~\ref{tab:heat_loads_sources_40K_table}, $\mathrm{HL}_\mathrm{1,~40-80K~circuit,~CB42-H1~Box} = 996.50~\mathrm{W}$ (Note that, the thermal shields are cooled exclusively  by either the 40KF or 80KR line. Therefore, unlike in the 4.5-5K circuit case, the heat load is not distributed between supply and return lines). Thus, we get
\begin{multline*}
h_\mathrm{1,~40KF,H1~Box} = \\ \frac{\mathrm{HL}_\mathrm{1,~40-80K~circuit,~CB42-H1~Box}}{\dot{m}_\mathrm{1,~40-80K~circuit,~total}} + h_\mathrm{supply,~40KF,~CB42} \\
= 263.21~\mathrm{J/g}.
\end{multline*}
The outlet enthalpy of the helium return from Exp.~4 to the H1 Box is $h_\mathrm{return1,~80KR,~H1~Box}@(\mathrm{60~K,~14~bar}) = 327.1~\mathrm{J/g}$, which, under the maximum boundary temperature condition, is equal to $h_\mathrm{return1,~80KR,~CB42}$. Hence, for all subsequent calculations, we use $h_\mathrm{return1,~80KR,~CB42}$ as the outlet enthalpy of experiments. The required mass flow rate for Exp.~4 using $\mathrm{HL}_\mathrm{1,~40-80K~circuit,~Exp.~4} = 500~\mathrm{W}$ and $\mathrm{HL}_\mathrm{1,~40-80K~circuit,~TL4} = 16 ~\mathrm{W}$ given in column 3 of Tab.~\ref{tab:heat_loads_sources_40K_table}, $h_\mathrm{1,~40KF,H1~Box}$, $h_\mathrm{return1,~80KR,~CB42}$, is then estimated as follows, $\dot{m}_\mathrm{1,~40-80K~circuit,~Exp.~4} = 8.08~\mathrm{g/s}$.

Next, the inlet enthalpy of the helium supply via the 40KF line of the ACPS to the experiments, $h_\mathrm{1,~40KF,~ACPS}$, is assumed to be equal to $h_\mathrm{1,~40KF,~H1~Box}$, since the heat load to the ACPS $\mathrm{HL}_\mathrm{1,~40-80K~circuit,~ACPS}$ is negligible compared to $\mathrm{HL}_\mathrm{1,~40-80K~circuit,~CB42-H1~Box}$. Using this assumption, we calculate the mass flow rate for Exp.~1 using $\mathrm{HL}_\mathrm{1,~40-80K~circuit,~Exp.~1} = 3700~\mathrm{W}$ and $\mathrm{HL}_\mathrm{1,~40-80K~circuit,~TL6} = 16~\mathrm{W}$ given in column 3 of Tab.~\ref{tab:heat_loads_sources_40K_table}, together with $h_\mathrm{1,~40KF,~ACPS}$ and $h_\mathrm{return1,~80KR,~CB42}$, and the result is $\dot{m}_\mathrm{1,~40-80K~circuit,~Exp.~1} = 58.51~\mathrm{g/s}$. In similar steps, the mass flow rates for the remaining experiments and cases are calculated and summarized in Tab.~\ref{tab:mass_flow_rates_40KF}.

\begin{table*}[htbp]
\centering
\caption{\label{tab:mass_flow_rates_40KF} Summary of calculated mass flow rates for different experiments shown in Fig.~\ref{fig_ACPS_HB_H1_40_80K_heat_load_v1} and their corresponding heat loads in Tab.~\ref{tab:heat_loads_sources_40K_table}.}
\begin{tabular}{p{3cm}p{2cm}p{2cm}p{2cm}p{2cm}p{2cm}p{2cm}p{2cm}}
\hline
Cases & $\dot{m}_\mathrm{i,~j,~total}$ (g/s) & $\dot{m}_\mathrm{i,~j,~Exp.~4}$ (g/s) & $\dot{m}_\mathrm{i,~j,~Exp.~1}$ (g/s) & $\dot{m}_\mathrm{i,~j,~Exp.~2}$ (g/s) & $\dot{m}_\mathrm{i,~j,~Exp.~3}$ (g/s)\\
\hline
1: 60 K and Scenario 1 & 70.12 & 8.08 & 58.51 & 2.60 & 0.72 \\
2: 60 K and Scenario 2 & 91.90 & 9.46 & 74.24 & 4.02 & 1.27 \\
3: 80 K and Scenario 1 & 30.10 & 4.24 & 25.05 & 1.11 & 0.31 \\
4: 80 K and Scenario 2 & 39.41 & 4.97 & 31.85 & 1.72 & 0.55 \\
\hline
\end{tabular}
\footnotetext{Here, $i = 1~-~4$ representing various cases as well as $j$ denoting 40-80K circuit.}
\end{table*}

\subsection{Final mass flow rate via various TLs of ACPS to the experiments}
After estimating the mass flows described above, it is necessary to include a certain contingency (i.e., close to 20\%) for the sizing of valves, process pipes, flow meters and other components in Fig.~\ref{fig_ACPS_Box_only_flowscheme}~and~\ref{fig_ACPS_H1_HB_flow_scheme}. Therefore, the final mass flows (or specific mass flows), along with the corresponding temperatures and pressures for various TLs to the experiments, are described in Tab.~\ref{tab:final_mass_flow_rates}. We have also considered experimental cases in which the heat load is determined in advance, but the experiments have not yet been developed. Therefore, these parameters could be used for setting-up future experiments.

\begin{table*}[htbp]
\centering
\caption{\label{tab:final_mass_flow_rates} Summary of final mass flow rates, temperatures, and pressures considered for getting the sizes of valves, pipes, and flowmeters for TL~1~-~3 to Exp.~1~-~3 and Exp.~4. (Note: the helium supply via the H1 box to Exp.~4 will have a slightly higher temperature between 6.1~K and 6.3~K due to the heat load to the HERA TL.)}
\begin{tabular}{p{6cm}p{3cm}p{3cm}p{3cm}}
\hline
Parameters & TL1 to Exp. 1 & TL2 and TL3 to Exp. 2 and Exp. 3 &  Exp. 4 \\
\hline
Mass flow rate range (g/s) for 4.5KF/5KR line & 6 - 30 & 2 - 10 & 3- 10 \\
Temperature range (K) for 4.5KF/5KR line & 4.50 - 6.13 & 4.50 - 6.13 & 5.40 - 6.13 \\
Pressure range (K) for the 4.5KF line & 2.5 - 3.5 & 2.5 - 3.5 & 2.5 - 3.5 \\
Pressure range (bar) for the 5KR line & 1.0 - 1.30 & 1.0 - 1.30 & 1.0 - 1.30 \\
Mass flow rate range (g/s) for 40KF/80KR line & 25 - 75 & 2 - 10 & 5 - 30 \\
Temperature range (K) for 40KF/80KR line & 45 - 80 & 45 - 80  & 45 - 80 \\
Pressure range (bar) for 40KF/80KR line & 13 - 14 & 13 - 14  & 13 - 14 \\
\hline
\end{tabular}
\end{table*}

\section{3D design, fabrication, factory acceptance test, site acceptance test of ACPS}
After determining the final mass flow rate range, temperature range, and pressure range for the various TLs to the experiments presented in Tab.~\ref{tab:final_mass_flow_rates}, we have defined the design parameters of the control and hand valves, flow meters, safety valves, pressure transmitters, subcooler bath, heaters, thermal shield, process pipes and others ACPS components (see them in Fig.~\ref{fig_ACPS_H1_HB_flow_scheme}) and discussed them in our ACPS specification document \cite{Schaffran2022}. This also includes the specific requirements for a sudden vacuum failure, quality control measures, and risk assessments as per European design \cite{Schaffran2022}, cross checked by DEMACO \cite{DEMACO2024}. Following this, the 3D design, fabrication, factory acceptance test (FAT), and warm site acceptance test (SAT) of the ACPS were carried out in cooperation with DEMACO \cite{DEMACO2024}.

\subsection{Cryogenic valves}
To determine the flow coefficient $K_{vs}$ of different control valves, in addition to the mass flow rate of helium supply and return, different operating temperatures, pressure conditions and pressure drops were considered. Additionally, they feature an equal-percentage (eq$\%$) opening characteristic (rangeability~1:100), while warm valves for mixing warm GHe have a linear opening characteristic. The calculated valve opening range was considered to be between 22 $\%$ and 70$\%$ with a corresponding pressure drop of 50–100 mbar. It was also taken into account that helium supply valves use nozzle/flap positioners while helium return valves have piezo positioners.

Most cryogenic valves are normally closed, except for those used during cooldown and warmup. These valves are normally open and can also be used to vent helium into the quench gas line during a power failure. The valves installed in the ACPS are mainly from WEKA company \cite{WEKA2024} and were selected from the WEKA datasheet based on availability for the required $K_{vs}$.

\subsection{Flowmeters}
A total of five Coriolis flowmeters are installed in the ACPS for measuring helium mass flow: four are connected to the 4.5KF lines, and one is connected to the 40KF line (see Fig.~\ref{fig_ACPS_Box_only_flowscheme} and Fig.~\ref{fig_HERA_TL_HB_H1_ACPS}). The pressure drop across them is considered to be less than 50 mbar under the assumption made for the operation parameters. Each flowmeter is installed on process pipes horizontally.

\subsection{Safety valves}
The orifice diameter of safety valves in the ACPS for both liquid and gas services was estimated by considering the dimensions of several key components, including the TL (located between ACPS and experiments), the sub-cooler bath, the PHEX, and internal process pipes. The heat flux to cryogenic components operating below 80 K was assumed to be 6000 W /m$^{2}$ in case of vacuum breakdown. It was also ensured according to the design code ISO 21013-3:2016 for sizing safety valves that the pressure drop in the process line should not exceed more than 3 $\%$ of the discharge pressure (which is set at 20 bar), as well as the discharge coefficient (CoD) was considered to be 0.58 for liquid service and 0.80 for gas service. Finally, the valves were procured from the company LESER.

\subsection{Subcooler bath with PHEX, process heater and He level sensor}
The heat transfer of the PHEX after estimating for both scenarios discussed in Sect. IV.A.$\emph{2}$ was specified with 500 W with a mass flow range of 3-50 g/s. The pressure drop was estimated to be less than 100 mbar for a length of 18 meters of copper tube. The capacity of the subcooling bath is about 500 liters of liquid helium. The subcooler bath is also equipped with two process heaters of 500 watts each and two superconducting level sensors, whereas one is acting as a spare (or redundant) device.

\subsection{Thermal shield} The thermal shield, which covers all components of the 4.5-5K circuit, is made of copper materials and has welded stainless steel tubes through which helium gas flows at a temperature in between 40 K and 50 K. The maximum temperature gradient on the shield is in the range of 5-6 K. It is also thermally connected to the heat sink plate of the valves operating below 6 K via a copper braid  at the appropriate height. In addition, the shield is attached to the vacuum jacket with chains, making heat transfer almost negligible and allowing for compensation of thermal shrinkage.

\subsection{Multi-layer insulation} To minimize the thermal radiation generated at 300 K, the cryogenic components in the ACPS, including thermal shields, cryogenic valves and process pipes operating in a temperature range of 40 K to 80 K are encased in 40 layers of multilayer insulation (MLI). Additionally, the cryogenic components operating at a temperature below 6 K facing the same radiation generated at 80 K is covered with 10 layers of MLI. With this number of MLIs, the thermal radiation could be limited to approximately 1 W/m$^{2}$ at 40 K and to about 0.1 W/m$^{2}$ below 6 K in both cases \cite{weisend2021cryostat}.

\subsection{Factory acceptance test} After the fabrication of ACPS at DEMACO, the FAT was carried out at both warm and liquid nitrogen temperature to ensure the specification requirements: (1)All internal process piping was subjected to a helium pressure test (with a mixture of helium and nitrogen) up to 28 bar according to the European Pressure Equipment Directive 2014/68/EU. (2) The movement of valves was tested within its mechanical operation range at both room and liquid nitrogen temperatures, without any visible, audible or noticeable jerks. (3) Each temperature sensor was checked by measuring their resistance values at room temperature and liquid nitrogen temperature. (4) Helium leak tests were performed at both room and liquid nitrogen temperatures and the leak rate was $< 1 \cdot 10^{-8} \ \text{mbar} \cdot \text{l/s}$ at both temperatures. After the FAT was passed successfully, the ACPS was transported to DESY.

\section{Installation of ACPS in the HNH}
\begin{figure*}[!hbt]
\centering
\includegraphics[clip, angle=0, width=1.0\linewidth]{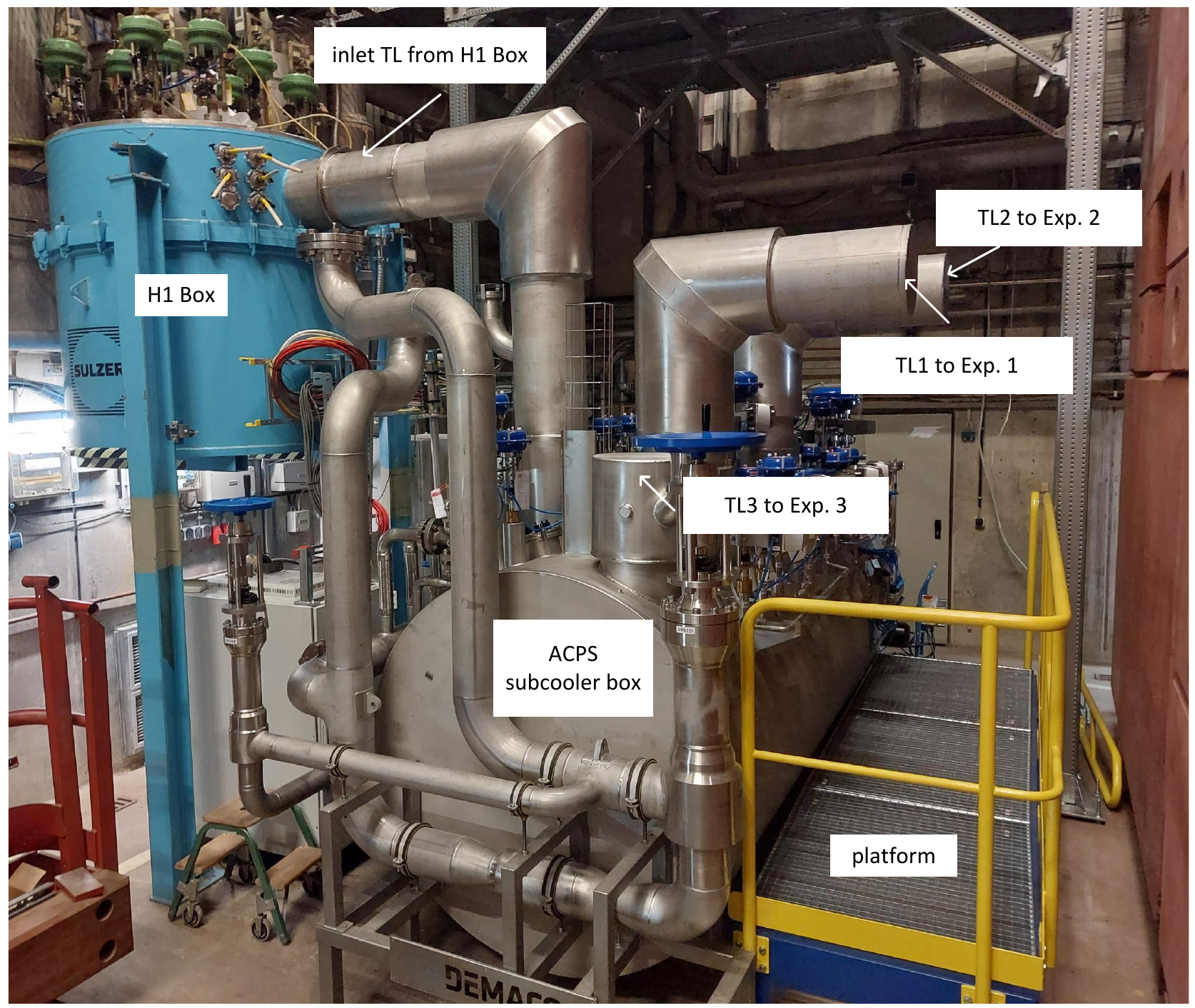}
\caption{ACPS photograph showing its current status.}
\label{fig_ACPS_Box_only}
\end{figure*}

After placing the ACPS in the HNH, the end caps of TLs to experiments were removed, and a short connection between forward and return lines was created, as shown in Fig. \ref{fig_ACPS_Box_only_flowscheme}. Later on all TL were covered with end caps along with safety valves. So now, Fig. \ref{fig_ACPS_Box_only} shows the current status of the ACPS and its connection via a transfer line to the H1 Box. It also displays TL1, TL2 and TL3 to the experiments.

\subsection{Interface between ACPS and H1 Box}
The interface between ACPS and H1 boxes was performed by adding a small piece of stainless steel pipes (which are in a length of 8-10 centimeters) and creating 8 welds, which were X-ray inspected, making sure that the joints were free of defects. The thermal shields of the H1 box and the ACPS are not thermally connected to each other, thus creating the space for the compensation of thermal shrinkage and the avoidance of thermal stresses, while both shields are covered with the MLI.

\subsection{Site acceptance test}
The SAT of the ACPS, like FAT, was performed at the HNH in the following steps: (1) The entire ACPS and all new interfaces (welds) were pressure tested to maximum 22 bar using with a mixture of gaseous helium and nitrogen and blocking SVs. (2) A successful completion of X-ray inspection and verification of few percentage of the total welds. (3) Helium leak tests on the ACPS were conducted by pumping the vacuum shell to a very high vacuum and then pressurizing process lines and subcooler bath with helium gas (as the internal parts of ACPS are not accessible). After that, a helium leak detector was used to monitor for signs of helium entering the vacuum shell. (4) The flange connections on the 300 K warm manifolds (which are not under vacuum) were tested by pressurizing the pipes and sniffing inside the bags covering the flanges. (5) Instrumentation tests were also carried out including checking the temperature sensors, pressure transmitters, process heaters, and smooth valve movement.

\subsection{Future plan}
In 2028, the ACPS is planned to be connected to the CB42 refrigerator in the DESY Cryo Plant reinstalling the TL between H1-Box and HERA-Box as shown in Fig.~\ref{fig_HERA_TL_HB_H1_ACPS}. Afterwards, we can conduct the SAT at nominal operating temperatures and pressures.

\section{Conclusion}
We have successfully completed the design, manufacturing, integration, warm and cold FAT, and warm SAT for the ACPS. We will reconnect the ACPS to the DESY Cryo Plant and then perform the cold SAT. Afterwards, we are ready to connect suitable experiments. This makes it a site capable of hosting a wide range of cryogenic experiments related to dark matter and gravitational-wave research.

\begin{acknowledgments}
We gratefully acknowledge the support provided by the Helmholtz Foundation for the Cryoplatform (ExNet\_0003). We also acknowledge support by the Deutsche Forschungsgemeinschaft (DFG, German Research Foundation) under Germany’s Excellence Strategy – EXC 2121 ``Quantum Universe'' – 390833306.\\
\end{acknowledgments}

\appendix
\section*{Appendix}
\section{Symbols}
We discuss the different notations used for the different quantities mentioned in the paper.
\begin{table*}[htbp]
\centering
\caption{\label{tab:Symbols} List of major symbols and their corresponding nomenclatures}
\begin{tabular}{p{3.5cm}p{10cm}}
\hline
\textbf{Symbol} & \textbf{Nomenclature} \\
\hline
$\mathrm{HL}_{\mathrm{1/2,\ 4.5-5K\ circuit,\ Exp.\ 1\text{-}4}}$ & Heat loads to the 4.5-5K circuit of Exp.~1–4 for Scenario~1 or 2. \\
$\mathrm{HL}_{\mathrm{1/2,\ 4.5-5K\ circuit,\ TL1\text{-}8}}$ & Heat loads to the 4.5-5K circuit of transfer lines TL1–8 for Scenario~1 or 2. \\
$\mathrm{HL}_{\mathrm{1/2,\ 4.5-5K\ circuit,\ total}}$ & Total heat load to the 4.5-5K circuit for Scenario~1 or 2. \\
$\mathrm{HL}_{\mathrm{1/2,\ 4.5KF,\ total}}$ & Total heat load to the 4.5KF line for Scenario~1 or 2. \\
$\mathrm{HL}_{\mathrm{1/2,\ 4.5KF,\ H1\text{-}Box}}$ & Heat load to the 4.5KF line up to the H1 Box location for Scenario~1 or 2. \\
$\mathrm{HL}_{\mathrm{1/2,\ 40-80K\ circuit,\ Exp.\ 1\text{-}4}}$ & Heat loads to the 40-80K circuit of Exp.~1–4 for Scenario~1 or 2. \\
$\mathrm{HL}_{\mathrm{1/2,\ 40-80K\ circuit,\ TL1\text{-}8}}$ & Heat loads to the 40-80K circuit of TL1–8 for Scenario~1 or 2. \\
$\mathrm{HL}_{\mathrm{1/2,\ 40-80K\ circuit,\ total}}$ & Total heat load to the 40-80K circuit for Scenario~1 or 2. \\

$h_{\mathrm{1/2,\ 4.5KF,\ H1\text{-}Box}}$ & Enthalpy of supercritical helium at the 4.5KF line at the H1 Box for Scenario~1 or 2). \\
$h_{\mathrm{1/2,\ 4.5KF,\ ACPS}}$ & Enthalpy of supercritical helium at the 4.5KF line at the ACPS for Scenario~1 or 2). \\
$h_{\mathrm{1/2,\ 40KF,\ H1\text{-}Box}}$ & Enthalpy of helium in the 40KF line at the H1 Box for Scenario~1 or 2). \\
$h_\mathrm{supply,~4.5KF,~CB42}$ & Enthalpy of helium supply from CB42 at 4.5 K and 3.5 bar. \\
$h_\mathrm{return,~5KR,~ACPS}$ & Enthalpy of helium return at ACPS at 4.5 K and 1.3 bar. \\

$\dot{m}_{\mathrm{1/2,\ 4.5KF,\ total}}$ & Total mass flow rate for the 4.5KF line at $\sim$4.5~K and 3.5~bar for Scenario~1 or 2). \\
$\dot{m}_{\mathrm{LR,\ Exps.\ 1~\text{-}~3}}$ & Helium liquefaction rate for Exp.~1~–~3 for Scenario 2\\
$\dot{m}_{\mathrm{1/2,\ 4.5KF,\ Exp.\ 1\text{-}4}}$ & Mass flow rate in the 4.5-5K circuit of Exp.~1~–~4 at $\sim$4.5~K and 3.5~bar for Scenario~1 or 2. \\
$\dot{m}_{\mathrm{1/2,~4.5KF,~subcooler}}$ & Evaporation rate of helium from the subcooler bath for Scenario~1 or 2 \\
$\dot{m}_{\mathrm{1/2,\ 40KF,\ total}}$ & Total mass flow rate in the 40KF line at $\sim$45~K and 14~bar for Scenario~1 or 2. \\
$\dot{m}_{\mathrm{1/2,\ 40KF,\ Exp.\ 1~\text{-}~4}}$ & Mass flow rate in the 40-80K circuit of Exp.~1~–~4 at $\sim$4.5~K and 3.5~bar for Scenario~1 or 2. \\

$\dot{Q}_{\mathrm{1/2,\ 4.5KF,\ PHEX}}$ & Heat transfer at the PHEX for Scenario~1 or 2. \\

$x_{\mathrm{1/2,\ 4.5KF,\ VC2110}}$ & Valve yield factor for JT valve VC2110 (Scenario~1 or 2). \\
$x_{\mathrm{1/2,\ 4.5KF,\ VC2111}}$ & Valve yield factor for JT valve VC2111 (Scenario~1 or 2). \\
\hline
\end{tabular}
\end{table*}

\section{A complete flow scheme of different cryogenic lines and systems}
\begin{figure*}[!hbt]
\centering
\includegraphics[angle=90, width=0.5\linewidth]{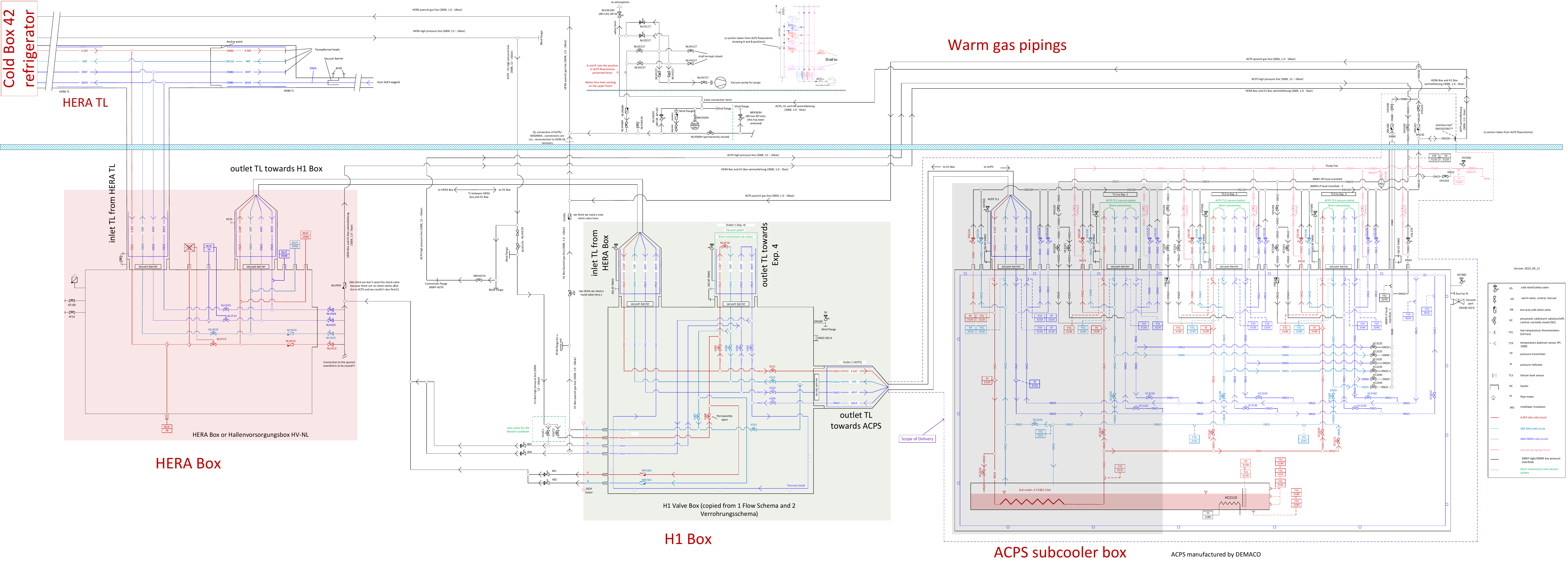}
\caption{A complete flow scheme of different cryogenic lines and systems including HERA Box, H1 Box and ACPS in the HNH area connected to CB42 refrigerator via HERA TL. Additionally, it shows warm gas line connections.}
\label{fig_ACPS_H1_HB_flow_scheme}
\end{figure*}

We provide a complete flowscheme highlighting HERA TL, HERA Box, H1 Box and ACPS in Fig. \ref{fig_ACPS_H1_HB_flow_scheme}. It also shows warm gas pipings of the ACPS and the H1 Box. A small section of it is presented in Fig. \ref{fig_ACPS_Box_only_flowscheme}.

\section{Yield factor calculation for JT Valve}
During expansion in the JT (Joule-Thomson) valve, helium is being produced into liquid and gaseous phases under constant specific enthalpy, there is no difference between the inlet enthalpy ($H_i$) and the outlet enthalpy ($H_f$). This phenomenon is referred to as isenthalpic expansion \cite{Jeong2018}. Thus, we can write:
\[
   H_i = H_o,
\]
or, in terms of mass flow rates:
\[
    \dot{m}_i h_i = \dot{m}_l h_l + \dot{m}_g h_g,
\]
where $\dot{m}_i$, $\dot{m}_l$, and $\dot{m}_g$ are the mass flow rates of inlet helium, oulet liquid helium, and outlet gaseous helium, respectively, and $h_i$, $h_l$, $h_g$ are their specific enthalpies. The mass balance is:
\[
    \dot{m}_i = \dot{m}_l + \dot{m}_g.
\]
Substituting this into the energy balance:
\[
    \dot{m}_i h_i = \dot{m}_l h_l + \left( \dot{m}_i - \dot{m}_l \right) h_g,
\]
and rearranging gives the fraction of mass flow liquefied:
\[
    \frac{\dot{m}_l}{\dot{m}_i} = \frac{h_i - h_g}{h_l - h_g}.
\]

This fraction, known as the yield factor, is:
\[
    x = \frac{\dot{m}_l}{\dot{m}_i}.
\]


\end{document}